# Conduction States with Vanishing Dimerization in Pt Nanowires on Ge(001) Observed with Scanning Tunneling Microscopy

J. Schäfer[1], D. Schrupp[2], M. Preisinger[2], and R. Claessen[1]

[1]*Physikalisches Institut, Universität Würzburg, 97074 Würzburg, Germany*
[2]*Institut für Physik, Universität Augsburg, 86135 Augsburg, Germany*

The low-energy electronic properties of one-dimensional nanowires formed by Pt atoms on Ge(001) are studied with scanning tunneling microscopy down to the millivolt-regime. The chain structure exhibits various dimerized elements at high tunneling bias, indicative of a substrate bonding origin rather than a charge density wave. Unexpectedly, this dimerization becomes vanishingly small when imaging energy windows close to the Fermi level with adequately low tunneling currents. Evenly spaced nanowire atoms emerge which are found to represent conduction states. Implications for the metallicity of the chains are discussed.

Solids with nearly one-dimensional (1D) properties represent a lower boundary for the miniaturization of electrical conductors, and let us expect particularly strong interactions of electrons and lattice owing to a reduction of electrostatic screening. Under these conditions, the 1D nature of the electron system can be responsible for the occurrence of a charge density wave (CDW), implying a metal-insulator transition for the electron band concerned [1]. Moreover, quasi-1D systems in the metallic phase offer an opportunity to study exotic physics that results from the predicted breakdown of the Fermi liquid picture, described as Luttinger liquid [2]. Challenging realizations of such 1D solids can be found in chains of adsorbates on surfaces which allow a particularly good control of the structural parameters. On the surface of semiconductors such as silicon and germanium, recently 1D reconstructions of metallic character, so-called *nanowires*, have been identified. Some of these nanowires, e.g. made of In atoms on Si(111) [3,4] or Au atoms on Si(553) [5,6], exhibit CDWs. However, while it has been observed that CDW nanowires can even exhibit precursor fluctuations above the transition temperature [7], it remains unresolved whether each nanowire system displays a phase transition at all.

A new and little explored class of nanowires are 1D reconstructions on the (001) surface of the tetrahedral semiconductors. On Si(001), nanowires have been reported for rare earth adatoms, e.g. Gd [8,9], which are believed to form silicide structures. Recently, the formation of noble metal nanowires was reported for the Ge(001) surface, using Pt adatoms [10,11] as well as Au adatoms [12]. This reconstruction forms upon deposition of a submonolayer coverage of noble metal at either elevated temperature or with subsequent annealing. The formation of nanowires by In atoms [13] on Ge(001) has also been reported. To date, however, the electronic properties of these systems on Ge(001) remain largely unclear. In particular, the Pt nanowire system was reported to be in a dimer-distorted phase that could be indicative of a CDW, yet no indication for a phase transition has been found in comparing scanning tunneling microscopy (STM) data at 300 K and 77K [11]. The conductivity data reported are in fact contradictory, providing reports of both metallic [10] and insulating behavior [11] for the Pt nanowires.

In this Paper we report on STM of Pt nanowires on the Ge(001) surface where particular attention is paid to the low energy properties, using a tunneling bias of down to 2 meV. Indications for CDW behavior, such as a dimerization of the metal adatoms, are investigated. STM images at high bias reveal dimers both along and sideways of the chains, consistent with structural support elements. Surprisingly, for low voltages corresponding to states near the Fermi level $E_F$, the dimerization is virtually lifted on the ridge of the nanowires. This observation relies on using adequately low tunneling currents. Spectroscopy data corroborate a finite conductivity. It thus emerges that the electron states near $E_F$ are essentially decoupled from the dimerized structure of the nanowire embankment.

Experimentally, Ge substrates (highly n-doped, resistivity < 0.4 Ohmcm) were prepared by repeated cycles of Ar sputtering (1 keV) and annealing to 800 ºC. This procedures results in large flat terraces of low defect density. Pt was evaporated using an electron beam evaporator. The sample received a post-deposition anneal at 600 ºC to allow formation of the well-ordered nanowire reconstruction. STM measurements were performed *in-situ* under ultra-high vacuum conditions at room temperature using an Omicron VT-STM apparatus.

The formation of Pt nanowires is seen in the STM data of Fig. 1(a). The length of the nanowires is usually terminated by terrace steps or defects. The distance between two nanowires is measured to be ~ 16 Å for most of them, while occasionally a spacing of ~ 24 Å is observed - an observation consistent with Gurlu et al. [10].

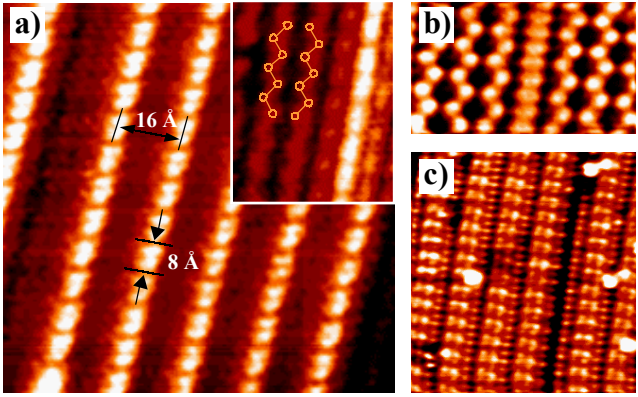

FIG. 1. a) STM image (90 Å × 90 Å) of Pt nanowires on the Ge(001) substrate, showing occupied states at −1.35 V. A *dimerization* along the nanowires with ~8 Å is evident. In the inset (to scale, -0.8 V, 0.5 nA) plain substrate is exposed, revealing a zigzag dimerization. b) On the *clean* Ge(001) surface (52 Å × 32 Å, −0.95 V, 0.5 nA), the (2×1) and c(4×2) phases coexist, the latter with an analogous zigzag spacing of ~8 Å. c) In Pt nanowires a dimerization is also present in the unoccupied states at +1.35 V (90 Å × 110 Å).

At a tunneling bias of −1.35 V as in Fig. 1(a), a *dimerization* along the chains is clearly observable in the topographic data. Along the wire, the corresponding repetition distance is ~ 8 Å. The structure between the nanowires can best be studied by looking at missing nanowires that expose the substrate, as in the inset to Fig. 1(a). Here a zigzag-like structure is observed, giving rise to a repetition period of also 8 Å.

This substrate structure may be compared to the clean Ge(001) surface prior to deposition in Fig. 1(b). A (2×1) and a c(4×2) reconstruction coexist [14,15] as a result of their approximate energetic equivalency. The atom spacing amounts to ~ 4 Å between dimers of the undistorted (2×1) phase. The repetition distance along the zigzag chains of the buckled c(4×2) phase is twice as much, ~ 8 Å. This buckling bears close resemblance to the substrate zigzag in the presence of nanowires, see inset to Fig. 1(a). Even if the free Ge surface were modified by dilute incorporation of Pt atoms [16], a direct relation to the dimer spacing of the adjacent nanowires exists, suggestive of a substrate-driven origin of the nanowire dimers.

In turning from occupied states to *unoccupied* states, a topography as shown in Fig. 1(c) is obtained. Here the prominent structure changes to *sideways* dimers. We note in passing that a left-right asymmetry is observed in the image, indicating that the building blocks for the nanowires are asymmetrically composed. A few defects may be due to surplus physisorbed atoms, yet they do not perturb the periodicity. The superstructure along the chain direction again measures ~ 8 Å, like in Fig. 1(a). From both measurements taken at high tunneling bias one must therefore conclude that, at least at high binding energies, a strong dimerized bonding exists, in registry with the underlying Ge surface lattice.

A dramatic change of the picture described thus far occurs when going to *lower bias*, as in the voltage sequence of Fig. 2. In the occupied states closer to the Fermi level at −250 mV in Fig. 2(a), the dimerization is largely suppressed. Although a weak "×2" superstructure can still be identified for most parts of the wires, the variation in the charge distribution along the nanowires has become marginal. In stark contrast to Fig. 1(a) at high bias, the overall impression of the nanowire ensemble is that they represent a spatially rather *uniform density of states* along the 1D direction, without interruption of the charge cloud between atoms.

In the STM image Fig. 2(b) showing unoccupied states at +200 mV, it is striking that the structure is now dominated by *single atom units* spaced ~ 4 Å. Dimers are no longer a prominent feature as in the images with high tunneling voltage in Fig. 1. Closer inspection of the image Fig. 2(b) still gives a faint indication of a dimer superstructure, yet this is limited to sideways protrusions on every second atom, rather than longitudinal dimers along the ridge of the nanowire.

It is intriguing to extend the analysis even closer to the Fermi level, where the actual conduction processes take place. When the tunneling voltage is lowered further as in Fig. 2(c) to V = +10 meV at 80 pA, the nanowires take the shape of a meandering line. An indication of u-shaped dimers is seen, which seems considerably stronger than in the preceding image Fig. 2(b) at intermediate bias. The u-shaped dimers exhibit a left-right asymmetry. In view of the corresponding asymmetry at high bias in Fig. 1(c), it seems plausible that this again relates to a contribution from the nanowire building blocks. It is important to note that this perception of *dimers* at the low bias of +10 meV is not a consistent continuation from the *monomer* image Fig. 2(b).

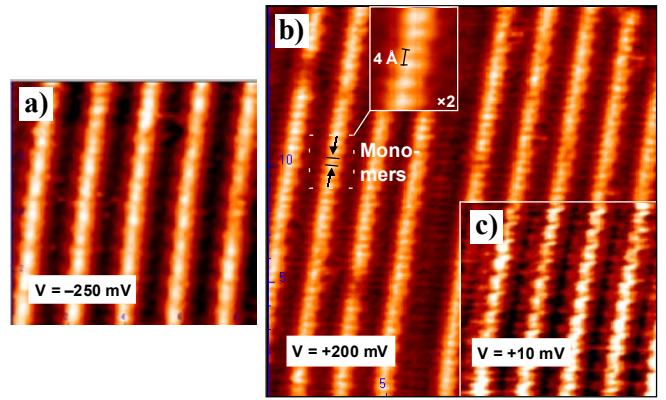

FIG. 2. STM images of Pt/Ge nanowires at moderate and low bias. a) Occupied states at -250 mV, 150 pA (88 Å × 88 Å), show only faint remainders of dimers. The local density of states is rather continuously distributed along the chains. b) Unoccupied states at + 200 mV, 150 pA (160 Å × 160 Å). The nanowire ridge is modulated by a *monomer* distance of ~4 Å. c) A low bias image with +10 mV, 80 pA exhibits a meandering pattern along the nanowires. Such high current load I/E leads to images with weak dimerization again.

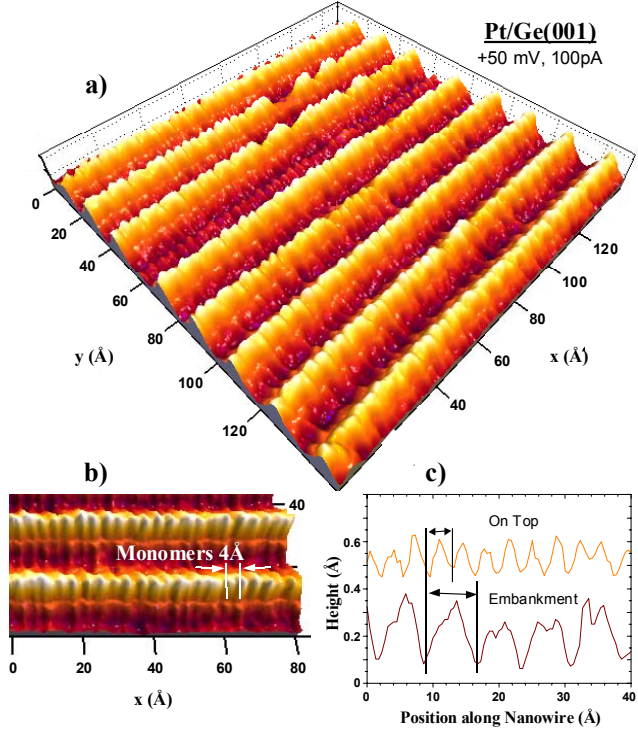

FIG. 3. STM images of Pt nanowires close to the Fermi level with low current load I/E. a) For +50 mV, 100pA, the dimerization is virtually absent. The current per binding energy I/E is 4× lower than in Fig. 3(c), implying a much larger tunneling distance. b) The enlarged view shows a pronounced monomer spacing of ~ 4 Å, rather than dimers. c) Line profiles for such images (curves offset for display) exhibit monomer spacings on the nanowire ridges, while along the embankment dimers become visible.

The *monomer* features can be recovered close to $E_F$ if the bias is somewhat larger, V = + 50 mV, at almost the same current, see Fig. 3(a). At T = 300 K the thermal broadening of the Fermi distribution amounts to 4 kT ≈ 100 meV, thus this dataset is still rather close to the conduction electron states. Here the charge distribution appears relatively even along the nanowires. In the image detail from the same data in Fig. 3(b), the monomers are clearly discernible, and they exhibit no obvious pairwise distortion.

Line profiles as in Fig. 3(c) show clearly that along the ridge of the nanowires, the dominant charge modulation is that of a monomer distance of ~ 4Å. Only when taking a profile along the embankment at about half height, i.e. closer to the groove between nanowires, a dimer modulation dominates. Noteworthy is also the exceptionally low variation in height along the ridge, which is less than 0.2 Å.

What can be the origin of these diverse observations about dimerization on the nanowires near $E_F$? In our STM data we find a consistent trend that the images strongly vary with the *tunneling current*. This current exponentially depends on the tunneling tip distance d, as evident from the simplified tunneling expression

$$I(V,d) \propto \int_{E_F-eV}^{E_F} \rho_S(\varepsilon+eV) \cdot \rho_{tip}(\varepsilon) \cdot e^{-d/d_0} d\varepsilon$$

where $\rho_s$ and $\rho_{tip}$ are the density of states of sample and tip, respectively, and $d_0$ is a constant given by the average tunneling barrier. Conversely, the tunneling probe distance is determined by the current *I* requested per binding energy window $E = e \cdot V$.

In our data series, the STM image Fig. 3(b) showing monomers was recorded at I/E = 100 pA / 50 meV = 2 pA/meV. In contrast, the image in Fig. 2(c) with weak dimerization was recorded at I/E = 8 pA/meV, which is 4× more and implying a much closer approach of the tip to the sample orbitals. The image with virtually complete absence of longitudinal dimers in Fig. 2(b) was taken with the particularly low current load of I/E = 0.75 pA/meV. Thus, we find that undimerized single units in the nanowires are seen at *low energies* and if the tip is *not closely approaching* the substrate. In turn, dimerization is present in the images at high binding energies and for close tip approaches detecting the backbonds into the substrate.

Concerning the question of metallicity, we have attempted to image the wires at lowest voltages. The result for –2 meV and only 20 pA current is shown in Fig. 4(a). Despite a current load of 10 pA/meV this image still does not show a prominent indication of nanowire dimerization, although we add that such low signal data do not lend itself to a detailed structural analysis. Rather, the case in point is that the structure of the nanowires can be imaged consistently down to these exceptionally low voltages. They exhibit a highly continuous local density of states along the 1D direction, which we take as one indication that they are *metallic* conductors.

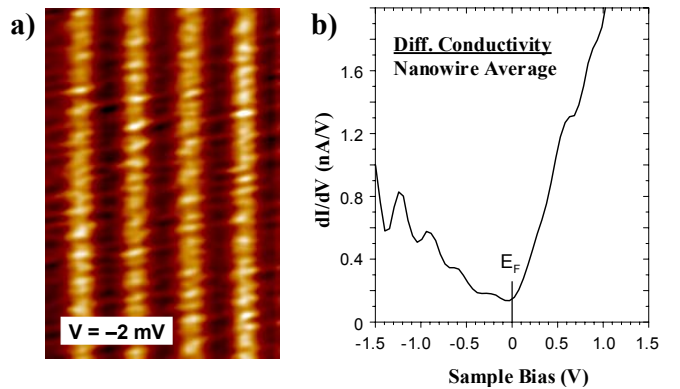

FIG. 4. a) STM data at –2 meV (70 Å × 100 Å) provide information about the vicinity of the Fermi level. An extremely low current of 20 pA results in a detection of monomeric states and a rather continuous 1D charge distribution. b) Spectroscopy data taken on the ridge of the nanowires (ensemble average), revealing a finite differential conductivity at $E_F$.

With respect to the conduction properties, we have also performed tunneling current spectroscopy, plotted as the differential conductivity dI/dV in Fig. 4(b). The data have been obtained from numerical differentiation of I-V-curves. This avoids the problem of phase shifts common in modulation measurements, which can be subject to varying space charges. The data is spatially averaged over the length extent of numerous nanowires. A finite conductivity at zero bias is found. A metallic conductivity was also reported in [10], while subsequent data by the same group [11] show hardly any conductivity at zero bias. Still the overall shape of the dI-dV-curve at 77 K [11] bears resemblance to ours. Since the substrate imposes a series resistance for the spectroscopy, we have taken care to use a sufficiently doped semiconductor, and this may explain the variance of [10] with [11]. A noteworthy related system on Si(001) are Gd nanowires. For these, metallic conductivity has been reported from photoemission [9] and is found to persist down to 20K, i.e. the chains are not exhibiting a CDW instability.

Our results strongly suggest that dimers are the building blocks of the nanowire embankment. That would be consistent with a dimerization period of ~ 8 Å already present in the Ge substrate. The dimer structure also noted by Gurlu *et al.* [10] is therefore likely due to the back-bonding of the nanowires. The electronic *conduction properties*, however, are determined very close to the Fermi level $E_F$, and here we find non-dimerized charge clouds reaching outside the nanowires. In considering the charge configuration of Ge, $4s^2\,4p^2$, it is conceivable that the extended charge clouds of the s- and p-orbitals are contributing to the intense and evenly distributed charge at –0.25 eV binding energy, see Fig. 2(a). For Pt, $4f^{14}\,5d^{10}\,6s^0$, the d-orbitals may donate some charge into the bonding with the Ge orbital, leading to a partially occupied, conducting d-band. Possibly the Pt s-orbitals are involved as well. However, the amount of charge transfer is not known, nor does a theoretical structure model currently exist. The sharp monomers observable in a window from approximately $E_F - 0.2$ eV to $E_F + 0.2$ eV could result from spatially well-defined and evenly spaced Pt d-orbitals that stick out of the chains. Their overlap will determine the conductivity in chain direction, leaving the fascinating question of possible spin-charge separation in a 1D atom chain.

In summary, a concise series of tunneling images over a wide range of unoccupied and occupied states was obtained, establishing dimer building blocks at high energies. Since dimer elements exist with sideways orientation, no obvious connection to a CDW can be made. Closer to the Fermi level by using a correspondingly low tunneling current, the dimerization appears lifted and nanowire atoms with monomer spacing become visible. Their metallic conductivity may arise from an independent electron band of the Pt atoms not involved in the dimer formation of the substrate backbonds. The indication of non-dimerized Fermi level states forming a 1D conduction path in these chains also spurs speculation about a conductivity that could relate to a Luttinger liquid. Further insight might be provided by investigations of a possible CDW instability of these electrons as a function of temperature, as well as angle-resolved photoemission concerning their band dispersion.

The authors are grateful to M. Wisniewski for technical support. Discussion of structural possibilities of nanowires with F. Bechstedt are gratefully acknowledged. This work was supported by the DFG (grant CL 124/3-2).

[1] G. Grüner, *Density Waves in Solids*, Addison-Wesley Publishing, Reading (1994).
[2] J. Voit, Rep. Prog. Phys. **57**, 977 (1994).
[3] J. Kraft, M. G. Ramsey, and F. P. Netzer, Phys. Rev. B **55**, 5384 (1997).
[4] I. G. Hill and A. B. McLean, Phys. Rev. B **56**, 15725 (1997).
[5] J. R. Ahn, P. G. Kang, K. D. Ryang and H. W. Yeom, Phys. Rev. Lett. **95**, 196402 (2005).
[6] P. C. Snijders, S. Rogge, and H. H. Weitering, Phys. Rev. Lett. **96**, 076801 (2006).
[7] H. W. Yeom, S. Takeda, E. Rotenberg, I. Matsuda, K. Horikoshi, J. Schäfer, C. M. Lee, S. D. Kevan, T. Ohta, T. Nagao, and S. Hasegawa, Phys. Rev. Lett. **82**, 4898 (1999).
[8] Y. Chen, D. A. A. Ohlberg, and R. S. Williams, J. Appl. Phys. **91**, 3213 (2002).
[9] H. W. Yeom, Y. K. Kim, E. Y. Lee, K.-D. Ryang, and P. G. Kang, Phys. Rev. Lett. **95**, 205504 (2005).
[10] O. Gurlu, O. A. O. Adam, H. J. W. Zandvliet, B. Poelsema, Appl. Phys. Lett. **83**, 4610 (2003).
[11] N. Oncel, A. van Houselt, J. Huijben, A. S. Hallback, O. Gurlu, H. J. W. Zandvliet, B. Poelsema, Phys. Rev. Lett. **95**, 116801 (2005).
[12] J. Wang, M. Li and E. I. Altman, Phys. Rev. B **70**, 233312 (2004).
[13] G. Falkenberg, O. Bunk, R. L. Johnson, J. A. Rodriguez, N. Takeuchi, Phys. Rev. B **66**, 035305 (2002).
[14] H. J. W. Zandvliet, Physics Reports **388**, 1 (2003).
[15] O. Gurlu, H. J. W. Zandvliet, and B. Poelsema, Phys. Rev. Lett. **93**, 066101 (2004).
[16] O. Gurlu, H. J. W. Zandvliet, B. Poelsema, S. Dag, and S. Ciraci, Phys. Rev. B **70**, 085312 (2004).